\documentclass[conference]{IEEEtran}
\IEEEoverridecommandlockouts
\usepackage{multirow}
\usepackage[dvips]{graphicx}
\usepackage{amsmath}
\usepackage[psamsfonts]{amssymb}
\usepackage{amsxtra}
\usepackage{threeparttable}

\usepackage{cite}
\usepackage{amsfonts}
\usepackage{algorithmic}
\usepackage{textcomp}
\usepackage{xcolor}
\usepackage{tabularx}
\usepackage{booktabs}

\usepackage{adjustbox}
\usepackage{float}

\def\BibTeX{{\rm B\kern-.05em{\sc i\kern-.025em b}\kern-.08em
    T\kern-.1667em\lower.7ex\hbox{E}\kern-.125emX}}
\begin{document}

\title{Speech Enhancement-assisted Voice Conversion in Noisy Environments}

\author{\IEEEauthorblockN{1\textsuperscript{st} Yun-Ju Chan}
\IEEEauthorblockA{\textit{Department of Electrical and Computer Engineering} \\
\textit{National Yang Ming Chiao Tung University}\\
Hsinchu, Taiwan}
\and
\IEEEauthorblockN{2\textsuperscript{nd} Chiang-Jen Peng}
\IEEEauthorblockA{\textit{Department of Electrical and Computer Engineering} \\
\textit{National Yang Ming Chiao Tung University}\\
Hsinchu, Taiwan}
\and
\IEEEauthorblockN{3\textsuperscript{rd} Syu-Siang Wang}
\IEEEauthorblockA{\textit{Department of Electrical and Computer Engineering} \\
\textit{Yuan Ze University}\\
Taoyuan, Taiwan}
\and
\IEEEauthorblockN{4\textsuperscript{th} Hsin-Min Wang}
\IEEEauthorblockA{\textit{Institute of Information Science} \\
\textit{Academia Sinica}\\
Taipei, Taiwan}
\and
\IEEEauthorblockN{5\textsuperscript{th} Yu Tsao}
\IEEEauthorblockA{\textit{Research Center for Information Technology Innovation} \\
\textit{Academia Sinica}\\
Taipei, Taiwan}
\and
\IEEEauthorblockN{6\textsuperscript{th} Tai-Shih Chi}
\IEEEauthorblockA{\textit{Department of Electrical and Computer Engineering} \\
\textit{National Yang Ming Chiao Tung University}\\
Hsinchu, Taiwan}
}

\maketitle

\begin{abstract}
Numerous voice conversion (VC) techniques have been proposed for the conversion of voices among different speakers. Although good quality of the converted speech can be observed when VC is applied in a clean environment, the quality degrades drastically when the system is run in noisy conditions. In order to address this issue, we propose a novel speech enhancement (SE)-assisted VC system that utilizes the SE techniques for signal pre-processing, where the VC and SE components are optimized in an joint training strategy with the aim to provide high-quality converted speech signals. We adopt a popular model, StarGAN, as the VC component and thus call the combined system as EStarGAN. We test the proposed EStarGAN system using a Mandarin speech corpus. The experimental results first verified the effectiveness of joint training strategy used in EStarGAN. Moreover, EStarGAN demonstrated performance robustness in various unseen noisy environments. The subjective listening test results further showed that EStarGAN can improve the sound quality of speech signals converted from noise-corrupted source utterances.
\end{abstract}


\section{Introduction}
    The goal of voice conversion (VC) is to convert a human voice from a source speaker into that of a predefined target, but preserve the linguistic content of the source speech \cite{toda2007voice,aihara2015many}. By applying a conversion function in terms of the speaker attributes to achieve a specific goal, numerous VC approaches have been proposed in the past, such as Gaussian mixture model (GMM) based methods \cite{toda2007voice,stylianou1998continuous} and exemplar-based techniques \cite{wu2014exemplar,wu2016locally}. Several studies leverage deep-learning-based techniques to promote VC to conduct non-linear transformations and synthesize high-quality speaker voices, which are much similar to the target voice \cite{desai2010spectral,sun2015voice,hsu2016voice,wu2017denoising}. For example, generative adversarial network (GAN) based approaches, such as CycleGAN \cite{kaneko2018cyclegan} and StarGAN \cite{kameoka2018stargan}, apply a generator composed of an encoder and a decoder to perform VC, while using an additional discriminator or classifier to handle the statistical distribution of the target speaker. 
    
    In general, the quality of the converted speech is reasonable in a clean background; however, the quality of VC applied in a noisy environment degrades dramatically. To address this issue, several works trained their systems directly under an interference background during the training stage; hence, robust VC could be achieved in online testing \cite{kurita2019robustness,miao2020noise,takashima2012exemplar,xie2021noisy}. For such VC systems with denoising function, a noisy source feature is placed at the input of the system; the target clean voice is then obtained at the output.
    
    Instead of deriving the VC model directly from the noisy--clean training pairs, we propose a novel approach that first enhances the source features through a speech enhancement (SE) model and then converts the enhanced features into the target speaker's voice. SE technology is an essential speech processing front-end that extracts clean components from the noise-corrupted input for downstream applications, such as hearing aids \cite{wang2017deep,lai2016deep}, automatic speech recognition \cite{weninger2015speech,li2014overview}, and speaker recognition \cite{kolboek2016speech,shon2019voiceid,a2015automatic}. Several SE approaches have also been incorporated with deep-learning-based neural network architectures to improve the system performance in non-stationary noisy environments \cite{xu2014regression,lu2013speech,pascual2017segan}. By applying a model that considers the contextual information from the input logarithmic power spectra (LPS), decent speech quality and intelligibility can be obtained \cite{weninger2015speech}.  
    
    We proposed a novel E-StarGAN VC system that integrates an SE function to handle a noisy background during VC tasks. The LSTM-based model structure was used to construct the SE function, whereas the following StarGAN architecture was used to conduct the conversion. AutoVC, which has been proven to provide a good VC performance in a clean environment, is used as the generator in the StarGAN. The experiment results clearly indicate that the newly proposed E-StarGAN VC model significantly improves the sound quality and similarity with the target speaker of the converted speech. In addition, E-StarGAN has proven to be a robust VC system under various noisy environments. In the future, we plan to study E-StarGAN for singing voice conversion. We investigated EStarGAN in terms of objective and subjective tests under different noisy conditions. The evaluation results suggest that when compared with the baseline AutoVC, the converted speech signals yield by EStarGAN, which jointly trained the SE and VC models, have lower Mel cepstral distortion (MCD) values and high signal quality and similarity. Moreover, by further optimizing the discriminator and classifier, we can obtain even better conversion performance.
    
    
    \section{Related work}\label{sec:relwork}
    The conventional StarGAN-VC model architecture is composed of generator $\mathbb{G}$, discriminator $\mathbb{D}$, and classifier $\mathbb{C}$ modules. The VC process is carried out through the generator, which is composed of an encoder and a decoder. The encoder is used to decompose a source acoustic feature $\mathbf{y}$ into a speaker-independent vector; it is then combined with the target speaker attribute $\mathbf{I}_t$ for the decoder to reconstruct the target feature $\mathbf{t}$. Notably, $\mathbf{I}_t$ is a one-hot vector, where the nonzero element corresponds to the target speaker identity. Therefore, we can formulate the VC process as $\mathbf{t}=\mathbb{G}\{\mathbf{y},\mathbf{I}_t\}$. The discriminator and classifier functions in the training stage are used as additional regularizers to perform StarGAN-VC. Specifically, adversarial, identity-mapping, speaker classification, and cycle-consistency losses \cite{kameoka2018stargan} are used and introduced in the following subsections.
    \subsection{Adversarial loss function}
    With respect to the discriminator and generator, the adversarial loss function is defined as
    \begin{equation}
        \begin{scriptsize}
            \begin{aligned}
                \mathcal{L}_{adv}^{\mathbb{D}}&=-{E}_{\mathbf{t}, \mathbf{I}_t}\{log[ \mathbb{D}(\mathbf{t}, \mathbf{I}_t)]\}-{E}_{\mathbf{y}, \mathbf{I}_t}\{log[1-\mathbb{D}(\mathbb{G}(\mathbf{y},\mathbf{I}_t),\mathbf{I}_t)]\},\\
                \mathcal{L}_{adv}^{\mathbb{G}}&=-{E}_{\mathbf{y}, \mathbf{I}_t}\{log[\mathbb{D}(\mathbb{G}(\mathbf{y},\mathbf{I}_t),\mathbf{I}_t)]\},
            \end{aligned}
        \end{scriptsize}
    \end{equation}
    where ${E}$ represents the expectation operation. We then minimize $\mathcal{L}_{adv}^{\mathbb{D}}$ for $\mathbb{D}$ to distinguish the real speech $\mathbf{t}$ from a fake sample $\mathbb{G}(\mathbf{y})$, and minimize $\mathcal{L}_{adv}^{\mathbb{G}}$ for $\mathbb{G}$ to generate a speech feature in the target domain that can deceive the discriminator. The discriminator output is a two-dimensional one-hot vector that indicates the true or false class of the associated input.
    
    \subsection{Speaker-classification loss function}\label{sec:cls}
    To further preserve the speaker characteristics from the generated speaker voice, classifier $\mathbb{C}$, which specifies the speaker identity from $K$ speakers is applied to the output side of the generator. Thus, the output dimension of $\mathbb{C}$ is $K$. The corresponding loss function is formulated as follows:
    \begin{equation}
        \begin{scriptsize}
            \begin{aligned}
                \mathcal{L}_{cls}^{\mathbb{C}}&=-{E}_{\mathbf{t}, \mathbf{I}_t}\{log[ p_{\mathbb{C}}(\mathbb{C}(\mathbf{t}))]\},\\
                \mathcal{L}_{cls}^{\mathbb{G}}&=-{E}_{\mathbf{y}, \mathbf{I}_t}\{log[ p_{\mathbb{C}}(\mathbb{C}(\mathbb{G}(\mathbf{y}, \mathbf{I}_t)))]\}.
            \end{aligned}
        \end{scriptsize}
    \end{equation}
    The loss is then minimized for the classifier to identify the input speaker and for the generator $\mathbb{G}$ to provide a voice that is similar to the target speaker.
    
    \subsection{Cycle-consistency loss function}
    In addition to applying the adversarial loss for StarGAN, the cycle-consistency loss was leveraged for a generator to preserve more acoustic content during the conversion process. The cycle-consistency loss function is defined as follows:
    \begin{equation}\label{eq:cyc}
        \begin{scriptsize}
            \begin{aligned}
                \mathcal{L}_{cyc}^{\mathbb{G}}={E}_{\mathbf{y}, \mathbf{I}_t, \mathbf{I}_s}\{\| \mathbb{G}(\mathbb{G}(\mathbf{y}, \mathbf{I}_t), \mathbf{I}_s)-\mathbf{y}\|_1\},
            \end{aligned}
        \end{scriptsize}
    \end{equation}
    where $\mathbf{I}_s$ represents the source speaker attribute. Thus, the generated output speech feature from the generator belongs to the source domain according to the provided source speaker information, $\mathbf{I}_s$.
    
    \subsection{Identity-mapping loss function}
    To further promote the generator for learning speech structures, the identity-mapping loss is used and defined in the following equation:
    \begin{equation}\label{eq:idm}
        \begin{scriptsize}
            \begin{aligned}
                \mathcal{L}_{idm}^{\mathbb{G}}={E}_{\mathbf{y}, \mathbf{I}_s}\{\| \mathbb{G}(\mathbf{y}, \mathbf{I}_s)-\mathbf{y}\|_1\}.
            \end{aligned}
        \end{scriptsize}
    \end{equation}
    From \eqref{eq:idm}, we expect the output of the generator $\mathbb{G}(\mathbf{y}, \mathbf{I}_s)$ to be identical to that of the model input $\mathbf{y}$.
    
    \begin{figure*}[!t]
        {\centering
            \includegraphics[width=0.9\textwidth]{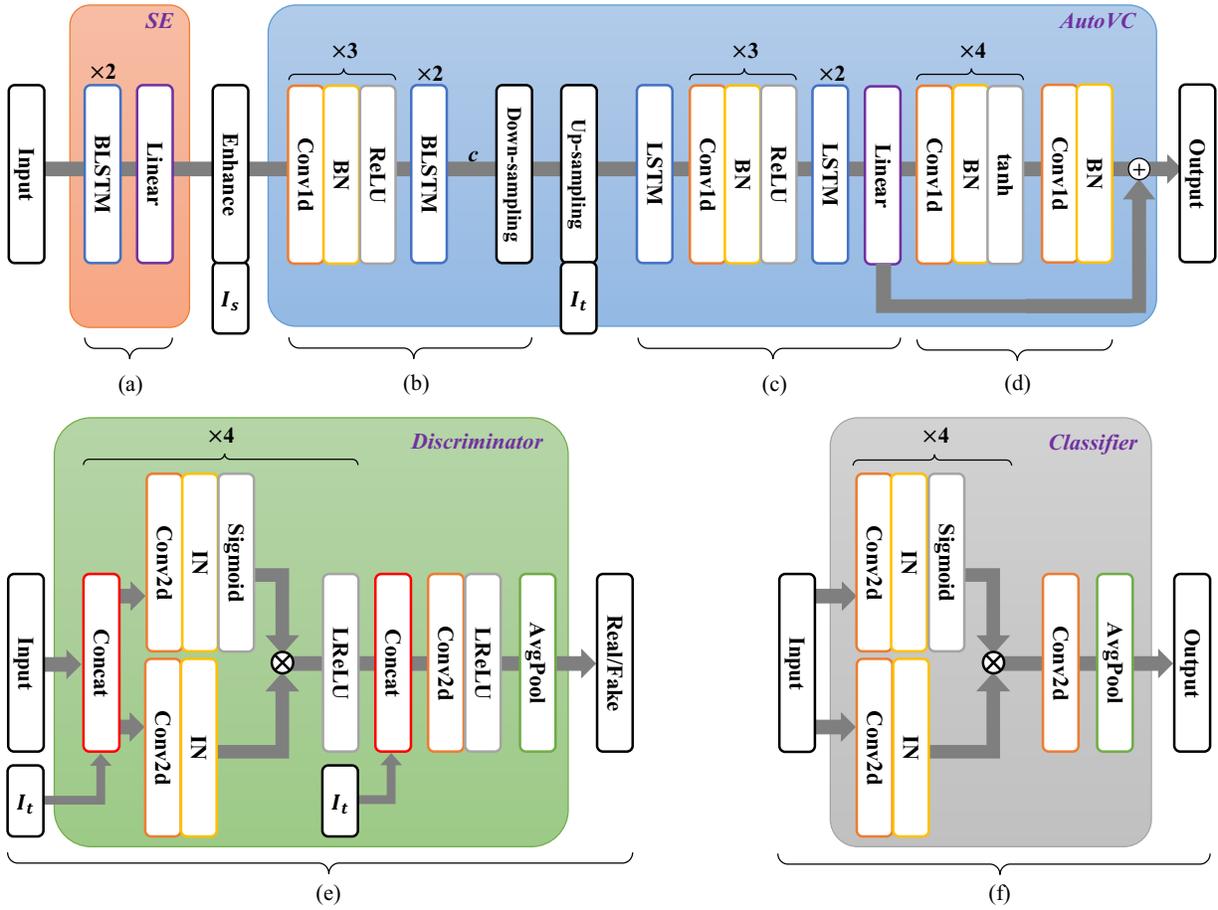}
            \caption{Block diagram of the proposed EStarGAN VC, which comprises six sub-systems: (a) SE, (b) encoder, (c) decoder, (d) post-net, (e) discriminator, and (f) speaker classifier. ``IN'' and ``BN'' represents the instance and batch normalization techniques, respectively \cite{qian2019autovc}. Notably, the upper row of the figure represents a block diagram of the generator (SE and AutoVC), wherein the input is a source noise feature, and the output is converted clean one.}\label{fig:frame}}
    \end{figure*}
    
    \section{The Proposed Approach}\label{sec:estar}
    In this study, we proposed EStarGAN for achieving VC in noisy environments. Similar to StarGAN, EStarGAN consists of a generator, discriminator, and speaker classifier. The detailed EStarGAN model is shown in Fig. \ref{fig:frame}, wherein the generator architecture is illustrated in the upper row, and the discriminator and speaker classifier are depicted at the bottom left and right, respectively. As mentioned in the Sec. \ref{sec:relwork}, the discriminator was applied to determine whether the input was true speech samples or the generated fake one, whereas the speaker classifier was employed to conduct the speaker identification task. Therefore, we focused on introducing the generator and training procedure used in this subsection.
    
    \subsection{Generator}
    From Fig. \ref{fig:frame}, the generator $\mathbb{F}$ is composed of SE and AutoVC components, where AutoVC comprises the encoder, decoder, and post-net processor. In addition, we formulated the VC process as $\hat{\mathbf{t}}=\mathbb{F}(\mathbf{x}, \mathbf{I}_s,\mathbf{I}_t)$ with respect to the noisy input, $\mathbf{x}$.
    
    \subsubsection{Feature enhancement}
    The BLSTM model was used to perform SE, which is composed of two BLSTM-hidden layers and followed by the 80-node feed-forward layer. For the noise-suppression task, we passed the noisy LMS $\mathbf{x}$ through the FE to provide an enhanced feature $\hat{\mathbf{y}}$ from the output of the SE model. The mean square error (MSE) $\mathcal{L}_{mse}\{\hat{\mathbf{y}},\mathbf{y}\}$ criterion was then applied to minimize the distance between $\hat{\mathbf{y}}$ and $\mathbf{y}$ for applying the BLSTM model during the training stage. In this study, 160 cells were used for each BLSTM hidden layer.
    
    \subsubsection{Auto-encoder based voice conversion}
    The block diagram of AutoVC is depicted in Fig. \ref{fig:frame}(b)--(d) and the conversion process is formulated as $\mathbb{F}_{auto}$. As the figure indicates, the encoder, $\mathbb{F}_{\mathbb{E}}$, encoded the enhanced LMS $\hat{\mathbf{y}}$ and the associated source speaker attribute $\mathbf{I}_s$ to generate the speaker-independent embedding feature, $\mathbf{c}=\mathbb{F}_{\mathbb{E}}(\hat{\mathbf{y}}, \mathbf{I}_s)$. This encoded feature comprised the outputs of the forward and backward paths in the latest BLSTM hidden layer. We then applied a 16-factor down-sampling followed by an up-sampling operation on the embedding feature along the time axis and then concatenated its outcome with the target speaker attribute $\mathbf{I}_t$ for the following decoder. The decoder output was the converted target LMS $\tilde{\mathbf{t}}$; this was filtered in the post-net thereafter to provide $\hat{\mathbf{t}}$ in the final output to shrink the possible signal distortions. Consequently, the relationship between input $\hat{\mathbf{y}}$ and output $\hat{\mathbf{t}}$ is $\hat{\mathbf{t}}, \tilde{\mathbf{t}}=\mathbb{F}_{auto}(\hat{\mathbf{y}}, \mathbf{I}_s,\mathbf{I}_t)$. In addition to $\hat{\mathbf{t}}$, $\hat{\mathcal{y}}, \tilde{\mathcal{y}}=\mathbb{F}_{auto}(\hat{\mathbf{y}}, \mathbf{I}_s,\mathbf{I}_s)$ was also used to conduct a loss function that was defined as:
    \begin{equation}\label{eq:auto}
        \begin{scriptsize}
            \begin{aligned}
                \mathcal{L}_{auto}=\mathcal{L}_{mse}\{\tilde{\mathcal{y}}, \hat{\mathbf{y}}\}+\mathcal{L}_{mse}\{\hat{\mathcal{y}}, \hat{\mathbf{y}}\}+\lambda_{auto}\mathcal{L}_{mse}\{\mathbb{F}_{\mathbb{E}}(\hat{\mathcal{y}}, \mathbf{I}_s), \mathbf{c}\}.
            \end{aligned}
        \end{scriptsize}
    \end{equation}
    Notably, $\mathbb{F}_{\mathbb{E}}(\hat{\mathcal{y}}, \mathbf{I}_s)$ in \eqref{eq:auto} represents the output $\hat{\mathcal{y}}$ was placed in the input side of $\mathbb{F}_{\mathbb{E}}$ to generate the associated speaker-independent embedding feature.
    \subsection{Training stage}
    \subsubsection{Loss functions}
    The loss functions, listed in \eqref{eq:estargan} were leveraged and minimized to perform EStarGAN VC.
    \begin{equation}\label{eq:estargan}
        \begin{scriptsize}
            \begin{aligned}
                \mathcal{L}^{\mathbb{F}}&=\mathcal{L}_{adv}^{\mathbb{F}}+\lambda_{cls}\mathcal{L}_{cls}^{\mathbb{F}}+\lambda_{cyc}\mathcal{L}_{cyc}^{\mathbb{F}}+\lambda_{idm}\mathcal{L}_{idm}^{\mathbb{F}},\\
                \mathcal{L}^{\mathbb{C}}&=\mathcal{L}_{cls}^{\mathbb{C}},\\
                \mathcal{L}^{\mathbb{D}}&=\mathcal{L}_{adv}^{\mathbb{D}}+\lambda_{gp}\mathcal{L}_{gp}^{\mathbb{D}},
            \end{aligned}
        \end{scriptsize}
    \end{equation}
    where
    \begin{equation}
        \begin{scriptsize}
            \begin{aligned}
                \mathcal{L}_{adv}^{\mathbb{D}}&=-{E}_{\mathbf{t}, \mathbf{I}_t}\{log[ \mathbb{D}(\mathbf{t},\mathbf{I}_t)]\}-{E}_{\mathbf{y}, \mathbf{I}_t}\{log[1-\mathbb{D}(\mathbb{F}(\mathbf{x}, \mathbf{I}_s,\mathbf{I}_t),\mathbf{I}_t)]\},\\
                \mathcal{L}_{adv}^{\mathbb{F}}&=-{E}_{\mathbf{x}, \mathbf{I}_s,\mathbf{I}_t}\{log[\mathbb{D}(\mathbb{F}(\mathbf{x}, \mathbf{I}_s,\mathbf{I}_t),\mathbf{I}_t)]\},\\
                \mathcal{L}_{cls}^{\mathbb{F}}&=-{E}_{\mathbf{x}, \mathbf{I}_s,\mathbf{I}_t}\{log[ p_{\mathbb{C}}(\mathbb{C}(\mathbb{F}(\mathbf{x}, \mathbf{I}_s,\mathbf{I}_t)))]\},\\
                \mathcal{L}_{cyc}^{\mathbb{F}}&={E}_{\mathbf{x}, \mathbf{I}_s,\mathbf{I}_t}\{\|\mathbb{F}_{auto}(\mathbb{F}(\mathbf{x}, \mathbf{I}_s,\mathbf{I}_t),\mathbf{I}_t, \mathbf{I}_s)-\mathbf{y}\|_1\},\\
                \mathcal{L}_{idm}^{\mathbb{F}}&={E}_{\mathbf{y}, \mathbf{I}_s}\{\| \mathbb{F}(\mathbf{x}, \mathbf{I}_s,\mathbf{I}_s)-\mathbf{y}\|_1\}.
            \end{aligned}
        \end{scriptsize}
    \end{equation}
    In addition, $\mathcal{L}_{gp}^{\mathbb{D}}$ proposed in \cite{gulrajani2017improved} was used as an additional regulizer for the discriminator in EStarGAN.
    
    \subsubsection{Training steps}
    The detailed training process consists of four steps: (1) We first extracted the LMS features for each clean and noise utterance. (2) By placing the noisy LMS features at the input, we pre-train the SE module for 220$k$ iterations in order to generate the corresponding clean features. (3) Next, we cascade the SE module and Auto-VC (encoder, decoder and post-net) module in Fig. \ref{fig:frame} and jointly train the two models based on Eq. \eqref{eq:auto} for 905$k$ training steps. (4) Finally, we fixed the SE module and train the discriminator, classifier, and Auto-VC iteratively based on Eq. \eqref{eq:estargan} for 380$k$ epochs.
    
    Notably, if we only implement the training steps (1), (2), and (3), we can also obtain a noise-robust VC system, termed ``\textit{jt}-SE+VC''. In this case, noisy LMS features in the online stage are processed by the concatenated and jointly trained SE+VC to generate the converted speech signals. For comparison, we perform a direct concatenation of SE and VC, which are trained individually and without joint-training. More specifically, the SE is trained using noisy-clean paired utterances, and the VC is trained using paired source-target clean utterances. Then, we directly cascade the SE and VC modules; hereafter, this direct concatenation setup is denoted as ``SE+VC''.
    
    \section{Experiments and analysis}
    \label{sec:exp}
    In the following subsections, we first introduce the experimental setup and then provide results along with discussions.
    \subsection{Experiment setup}
    The Taiwan Mandarin Hearing in Noise Test \cite{huang2005development} script, which contains 320 sentences was used to form our database. The training and testing scripts were pronounced by four male and four female Mandarin speakers ($K=8$ in Sec. \ref{sec:cls}) and recorded at a 16 kHz sampling rate in a quiet meeting space to provide 2,560 utterances. We selected 310 samples per speaker and prepared 2,480 clean utterances to provide the training set among these recordings. Each training waveform was artificially deteriorated by 100 different types of noises \cite{100noise} signal-to-noise ratios (SNRs) of $15$, $10$, and $5$ dB to finally provide 744,000 noisy--clean pairs for model training. Meanwhile, the other 80 clean recordings were contaminated by ``engine,'' ``pink,'' ``street,'' and ``white'' noises, which selected from \cite{varga1993assessment} and freesound\footnote{https://freesound.org/}, at three SNRs of $5$, $10$, and $15$ dB to generate the corresponded noisy testing data. As a result, 960 noisy utterances were used for testing.
    
    We applied a short-time Fourier transformation with a 64-ms frame size and a 16-ms hop length to decompose a waveform for the feature extraction process. The prepared spectral features were then filtered using mel filters and generated 80-dimensional LMS for this study. For a waveform synthesis from the LMS, we used the parallel WaveGAN model proposed in \cite{yamamoto2020parallel}. The model was pre-trained using the COSPRO corpus \cite{tseng2005sinica} and then retrained on the training database described in this study. The detailed EStarGAN model architecture was illustrated in Fig. \ref{fig:frame}. For the one-dimensional convolutional layer (Conv1d) used in the generator of EStarGAN, the channel, kernel, and stride sizes were 512, 5, and 1, respectively. The rest configuration was introduced as follows:
    \begin{itemize}
        \item \textbf{Encoder}: The cell size of each of two LSTM hidden layers was 512.
        \item \textbf{Decoder}: The 512, 1024, and 1024 cells were sequentially used to construct hidden LSTM layers. The output layer was an 80-node feed-forward network.
        \item \textbf{Post-net}: The latest output feed-forward layer contained 80 nodes.
        \item \textbf{Discriminator}: Five two-dimensional convolutional (Conv2d) layers were stacked in order of [32, (3,9), (1,1)], [32, (3,8), (1,2)], [32, (3,8), (1,2)], [32, (3,6), (1,2)] and [40, (80,5), (1,1)], which follow the setting representation ``[channel size, kernel size, stride size]''. The dimension of the final output from the applied average pooling (AvgPool) process was one.
        \item \textbf{Speaker classifier}: Five Conv2d layers were used in order of [8, (2,2), (4,4)], [16, (2,2), (4,4)], [32, (2,2), (4,4)], [16, (1,2), (5,4)] and [8, (1,2), (6,4)]. The AvgPool operation finally provided an eight-dimensional output.
    \end{itemize}
    
    The model configuration of AutoVC was identical to that used in the EStarGAN generator. Notably, despite the prepared parallel database, the training process for AutoVC and EStarGAN was conducted under non-parallel conditions. Finally, we applied the MCD objective metric to evaluate the proposed system.
    
    \begin{figure}[t]
        \centering
        \centerline{\includegraphics[width=\columnwidth]{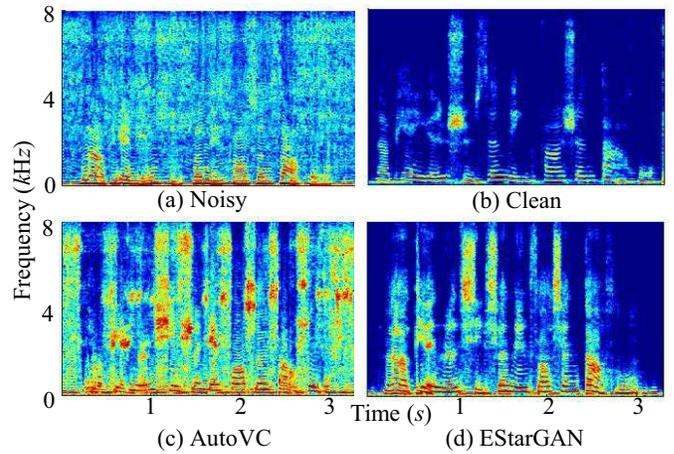}}
        \caption{The spectrograms of (a) Source noisy, (b) Target clean and the source noisy speech converted by (c) AutoVC and (d) EStarGAN.}\label{fig:spec}
    \end{figure}
    
    \subsection{Experiment results}
    To visualize VC performances, we present the LPS of a single source noisy utterance in Fig. \ref{fig:spec}(a), as well as the clean target speech in Fig. \ref{fig:spec}(b). The noisy utterances converted by AutoVC and EStarGAN are depicted in Figs. \ref{fig:spec}(c) and (d), respectively. In addition, the noisy source and clean target voices were pronounced by two different speakers. Comparing Figs. \ref{fig:spec}(a)(b)(c), the AutoVC spectrogram shows a clear acoustic structure in the voice parts, while introducing additional noise components in the silence segments. By contrast, EStarGAN converted speech in Fig. \ref{fig:spec}(d) shows a cleaner LPS and more detailed harmonic structures than those in \ref{fig:spec}(c).
    
    \begin{table}[!b]
        \caption{Averaged MCD values of AutoVC, SE+VC, \textit{jt}-SE+VC and EStarGAN in M2M, F2F, M2F and F2M conditions.}\label{tab:mcd}
        \begin{center}
            \begin{tabularx}{\columnwidth}{>{\raggedleft}m{1.5cm}>{\centering}m{1cm}>{\centering}m{1cm}>{\centering}m{1cm}>{\centering}m{1cm}>{\centering\arraybackslash}X}
                \toprule
                & \textbf{M2M} & \textbf{F2F} & \textbf{M2F} & \textbf{F2M} & \textbf{Avg.}\\
                \hline
                \textbf{AutoVC} & 9.43 & 9.70 & 10.03 & 9.48 & 9.66\\
                \textbf{SE+VC} &9.45&9.71&9.75&9.56&9.62\\
                \textbf{\textit{jt}-SE+VC}&7.74&7.95&8.17&8.35&8.05\\
                \textbf{EStarGAN} & \textbf{7.48} & \textbf{7.83} & \textbf{7.92} & \textbf{8.17} & \textbf{7.85}\\
                \bottomrule
            \end{tabularx}
        \end{center}
    \end{table}
    \begin{table}[!b]
        \caption{Averaged MCD values of AutoVC, SE+VC, \textit{jt}-SE+VC and EStarGAN in engine, pink, white, street noise environments.}\label{tab:mcdNsy}
        \begin{center}
            \begin{tabularx}{\columnwidth}{>{\raggedleft}m{1.5cm}>{\centering}m{1.4cm}>{\centering}m{1.4cm}>{\centering}m{1.4cm}>{\centering\arraybackslash}X}
                \toprule
                & \textbf{engine} & \textbf{pink} & \textbf{white} & \textbf{street}\\
                \hline
                \textbf{AutoVC} & 9.55 & 9.66 & 9.79 & 9.63\\
                \textbf{SE+VC} &9.61&9.62&9.63&9.62\\
                \textbf{\textit{jt}-SE+VC}&8.03&8.05&8.08&8.05\\
                \textbf{EStarGAN} & \textbf{7.83} & \textbf{7.84} & \textbf{7.88} & \textbf{7.86}\\
                \bottomrule
            \end{tabularx}
        \end{center}
    \end{table}
    \begin{table}[!b]
        \caption{Averaged MOS scores of AutoVC, SE+VC, \textit{jt}-SE+VC and EStarGAN in white, street noise environments.}\label{tab:mos}
        \begin{center}
            \begin{tabularx}{\columnwidth}{>{\raggedleft}m{2cm}>{\centering}m{1.9cm}>{\centering}m{1.9cm}>{\centering\arraybackslash}X}
                \toprule
                & \textbf{white} & \textbf{street} & \textbf{Avg.}\\
                \hline
                \textbf{AutoVC} & 1.91 & 1.85 & 1.88\\
                \textbf{EStarGAN} & \textbf{3.66} & \textbf{3.64} & \textbf{3.65}\\
                \bottomrule
            \end{tabularx}
        \end{center}
    \end{table}
    \begin{table}[!b]
        \caption{Averaged SIM values of AutoVC, SE+VC, \textit{jt}-SE+VC and EStarGAN in white, street noise environments.}\label{tab:sim}
        \begin{center}
            \begin{tabularx}{\columnwidth}{>{\raggedleft}m{2cm}>{\centering}m{1.9cm}>{\centering}m{1.9cm}>{\centering\arraybackslash}X}
                \toprule
                & \textbf{white} & \textbf{street} & \textbf{Avg.}\\
                \hline
                \textbf{AutoVC} & 2.63 & 2.85 & 2.73\\
                \textbf{EStarGAN} & \textbf{1.94} & \textbf{1.98} & \textbf{1.96}\\
                \bottomrule
            \end{tabularx}
        \end{center}
    \end{table}
    
    We evaluated the conversion performances of (a) male to male (M2M), (b) male to female (M2F), (c) female to male (F2M), and (d) female to female (F2F) in terms of the MCD metric, as shown in Table. \ref{tab:mcd}. For this comparison, the average MCD values between the EStarGAN-converted source speech and the target ground truth were denoted as ``EStarGAN''. The scores between the AutoVC-converted voices and the target speech were represented as ``AutoVC''. In addition, we listed the MCD results of SE+VC and \textit{jt}-SE+VC in the table for more comparisons. The average result among all testing utterances for each VC system is also listed in the table (denoted as ``Avg.'').
    
    From the table, SE+VC (directly concatenating SE and VC modules without joint-training) slightly improve the performance of AutoVC in terms of the Avg value. This observation suggests that the enhanced source feature cannot bring clear benefits for the following AutoVC-based StarGAN. One possible explanation is that the statistical properties of an enhanced speech features do not match well with the following VC generator because of the independent training processes for SE and VC. Contrarily, both \textit{jt}-SE+VC and EStarGAN consistently yield superior results to the baseline AutoVC and SE+VC under all testing conditions, as well as for Avg. The results demonstrate that the advantages of the joint-training strategy. Moreover, EStarGAN outperforms \textit{jt}-SE+VC,suggesting that  further improvement can be attained by performing joint-training on the discriminator and classifier in addition to the AutoVC generator.
    
    \begin{figure}[t]
        \centering
        \centerline{\includegraphics[width=\columnwidth]{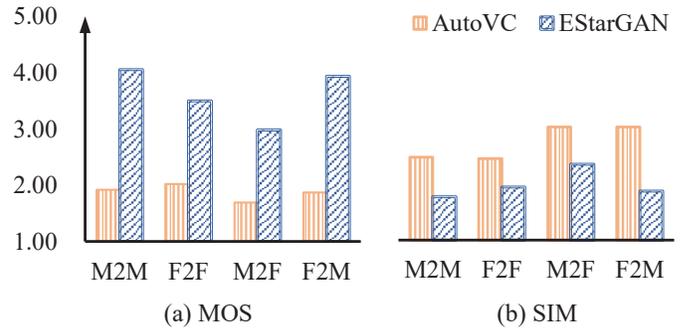}}
        \caption{The (a) MOS score and (b) SIM value for AutoVC and EStarGAN during subjective tests.}
        \label{fig:mossim}
    \end{figure}
    
    Next, we evaluate AutoVC, SE+VC, \textit{jt}-SE+VC and EStarGAN in terms of the average MCD values under four noisy environments. The results were showed in Table \ref{tab:mcdNsy}. Notably, the tested noisy conditions differed from those in the model training stage. From the table, EStarGAN provides the lowest MCD values for all testing noisy conditions when compared to other VC approaches. The result confirms the effectiveness of the joint training strategy and demonstrates the robustness of the proposed EStarGAN in performing noisy speech conversion.
    
    In addition to the objective evaluations, we also conducted subjective listening tests in terms of the mean opinion score (MOS) and similarity (SIM) for AutoVC and EStarGAN. The value range of each MOS and SIM was $[1,5]$. A higher MOS score indicates a better sound quality. However, a lower SIM value leads to a higher similarity. A single-blind listening test was conducted on 15 untrained but experienced normal hearing subjects who did not know which VC system was used. Two audio clips comprising a converted speech and a reference sound were presented for each participant. They were then asked to give an MOS score for the converted speech and the SIM value to determine the level of similarity for the heard samples. Listening samples were selected from the testing database and prepared in the following processes. The source utterances pronounced by two male and two female speakers were contaminated by white and street noises at an SNR of 10 dB and then converted into the clean target speech, which was uttered by all eight speakers. From these voice samples, we selected 40 EStarGAN and 40 AutoVC converted voices to conduct this subjective test. The averaged MOS and SIM with respect to the four gender-based conditions are illustrated in Fig. \ref{fig:mossim}. The results of MOS and SIM with respect to both noises are listed in Tables \ref{tab:mos} and \ref{tab:sim}, respectively.
    
    From Fig. \ref{fig:mossim}, both MOS and SIM values of EStarGAN outperform those of AutoVC under all evaluated conditions. These results demonstrate that the proposed EStarGAN system can provide reasonable sound quality and similarity for converting both male and female speakers. In addition, from Tables \ref{tab:mos} and \ref{tab:sim}, it can be seen that EStarGAN again provides better MOS and SIM values than AutoVC, revealing that EStarGAN can provide superior conversion effectiveness for human listeners in noisy environments.
    
    \section{Conclusions}\label{sec:con}
    We proposed a novel EStarGAN VC system that integrates an SE function to handle a noisy background during VC tasks. The LSTM-based model structure was used to construct the SE function, whereas the following StarGAN architecture was used to conduct the conversion. AutoVC, which has been proven to provide a good VC performance in a clean environment, is used as the generator in the StarGAN. The experiment results clearly indicate that the newly proposed EStarGAN VC model significantly improves the sound quality and similarity with the target speaker of the converted speech. In addition, EStarGAN has proven to be a robust VC system under various noisy environments. In the future, we plan to study EStarGAN for singing voice conversion.
    
    \bibliographystyle{ieeetr}
    \bibliography{refs}

\end{document}